\documentclass[a4paper,fleqn,usenatbib]{mnras}

\usepackage{newtxtext,newtxmath}

\usepackage[T1]{fontenc}
\usepackage{ae,aecompl}

\usepackage{graphicx}	
\usepackage{amsmath}	
\usepackage{amssymb}	

\usepackage{xcolor}

\title[Satellite galaxies in the Illustris-1 simulation]{Satellite galaxies in 
the Illustris-1 simulation: anisotropic locations around relatively isolated
hosts}

\author[T. G. Brainerd and M. Yamamoto]{
Tereasa G. Brainerd\thanks{E-mail: brainerd@bu.edu (TGB)}
and Masaya Yamamoto
\\
Boston University, Department of Astronomy, 725 Commonwealth Ave., Boston, MA, USA 02215\\
}

\date{Accepted XXX. Received YYY; in original form ZZZ}

\pubyear{2018}

\begin{document}
\label{firstpage}
\pagerange{\pageref{firstpage}--\pageref{lastpage}}
\maketitle

\begin{abstract}
We investigate the locations of satellite galaxies in the
$z = 0$ redshift slice of the hydrodynamical Illustris-1 simulation.
As expected from previous work,
the satellites are distributed anisotropically in the plane of the
sky, with a preference for being located near the major axes of their 
hosts.  Due to misalignment of mass and light within the hosts,
the degree of anisotropy 
is considerably less when satellite locations are
measured with respect to the hosts' stellar surface mass density than
when they are measured with respect to the hosts' dark matter surface mass density.  
When measured with respect to the hosts' dark matter surface mass
density, the mean satellite location depends strongly on host stellar
mass and luminosity, with the satellites of the faintest,
least massive hosts showing the greatest anisotropy.
When measured with respect to the hosts' stellar surface mass
density, the mean satellite location is essentially independent of
host stellar mass and luminosity.  In addition, the satellite
locations are largely insensitive to the amount of stellar mass used to
define the hosts' stellar surface mass density, as long as at least
50\% to 70\% of the hosts' total stellar mass is used.
The satellite locations
are dependent upon the stellar masses of the satellites, 
with the most massive satellites having the most anisotropic 
distributions.
\end{abstract}





\section{Introduction}

Small, faint satellite galaxies, in orbit around large, bright
`host' galaxies, have the potential to yield information about the
relationship between dark and luminous matter within the host galaxies.  In
particular, the spatial distribution of satellite galaxies may yield
constraints on the
shapes of the hosts' dark matter haloes and on the orientations of the 
hosts within those haloes.  Studies of host-satellite samples selected
from large redshift surveys such as the Two Degree Field Galaxy 
Redshift Survey (2dFGRS; Colless et al.\ 2001, 2003) and the Sloan Digital Sky
Survey (SDSS; Fukugita et al.\ 1996;
Hogg et al.\ 2001; Smith et al.\ 2002; Strauss et al.\ 2002;
York et al.\ 2000) have shown that, in the case of relatively
isolated host galaxies, the satellites exhibit a spatial anisotropy with respect
to the luminous major axes of their hosts.

By `relatively isolated', we mean host-satellite systems that were obtained
using a set of criteria that are intended to select systems in which the
gravitational potential is dominated by a single, bright galaxy and which
are meant to
exclude galaxy groups or clusters.  In the observed universe, the
isolation criteria are such
that both the Milky Way and M31, if viewed by an external observer, should be excluded
from the sample.  Due to the fact that line-of-sight distances to 
candidate host and satellite galaxies are not generally known in 
an observational sample, it is not possible
to select observed host-satellite systems using proximity in 3-dimensional space.  Hence,
the isolation criteria rely on line-of-sight velocity differences, which introduces
some level of uncertainty.  Most host-satellite systems obtained with
these selection criteria are somewhat more isolated than are the Milky Way or M31 (i.e.,
for the most part, systems like the two large subcomponents of the Local Group are rejected,
although often just barely; see \'Ag\'ustsson 2012).  However,
some less isolated systems, including loose groups, inevitably wind up being included 
in observational samples.  Because of this, we acknowledge that not all of the
systems are as well isolated as would be truly desired and, hence, we consider it more
appropriate to refer to these systems as `relatively isolated' rather than `isolated'.

When averaged over 
all satellites and all hosts, the satellites of relatively isolated host galaxies are
found preferentially close to the hosts' major axes 
(Sales \& Lambas 2004, 2009; Brainerd 2005; Azzaro et al.\ 2007; Bailin et al.\ 2008; 
\'Ag\'ustsson \& Brainerd 2010, 2011).
The degree to which observed satellite galaxies exhibit a spatial anisotropy 
has also been shown to be strongly dependent upon the physical properties
of the hosts and satellites
(Sales \& Lambas 2004, 2009; Azzaro et al.\ 2007; Bailin et al.\ 2008;
\'Ag\'ustsson \& Brainerd 2010, 2011).  In host-satellite samples obtained
from large redshift surveys of the observed universe, the most pronounced tendency for 
satellites to be found near their hosts' major axes occurs for hosts that
have high stellar masses, low star formation rates and are
red in colour.  Further, satellites that
have high stellar masses, low star formation rates and are red in colour show
a much greater spatial anisotropy than do satellites 
that have low stellar masses, high star formation rates and are blue
in colour.

Conclusions regarding the locations of satellite galaxies with respect 
to the disks of late-type 
host galaxies remain murky at present.  In a study of the satellites of
seven low-redshift spiral galaxies, Yegorova et al.\ (2011) found the angular
distribution of the satellites to be isotropic with respect to the 
position angles of their host galaxies.  In a study of the
satellites of massive galaxies with redshifts $0.1 \le z \le 0.8$, Nierenberg
et al.\ (2012) also found the satellites of late-type hosts to be
distributed isotropically with respect to the orientations of their
host galaxies.  Within the Local Group, there is evidence suggesting that 
a significant fraction of the satellites of M31 and the Milky Way are
distributed anisotropically, but with a preference for `polar' alignment
of some types of satellites
(e.g., Koch \& Grebel 2006; Metz et al. 2007, 2009).

Since Cold Dark Matter (CDM) haloes are triaxial (e.g., Jing \& Suto 2002),
an anisotropy in the satellite locations is expected 
if satellites are fair tracers of their hosts' dark matter haloes.  
For example, 
Zenter et al.\ (2006) found that the spatial distribution
of subhaloes in a `dark matter only' CDM simulation were distributed
anisotropically, with a preference for alignment with the major axes
of the central hosts' dark matter haloes.  In addition, in a study 
of group-mass systems in an N-body simulation, Faltenbacher et al. (2008) 
found that the satellites were located preferentially along the 
major axes of the central substructures of the groups.

If,
in projection on the sky, luminous host galaxies are well-aligned with
the mass density of their triaxial CDM haloes, one would
expect the satellite galaxies to be distributed anisotropically relative
to the luminous major axes of the hosts.
\'Ag\'ustsson \& Brainerd (2006; hereafter AB06) investigated the locations of satellite galaxies 
in the $z = 0$ redshift slice of the $\Lambda$CDM GIF simulation
(Kauffmann et al.\ 1999), which combined
a semi-analytic galaxy formation model (`SAM') with an adaptive 
particle-particle-particle-mesh N-body simulation. 
AB06 found that, if mass and light were perfectly
aligned within the hosts, the degree of anisotropy in the satellite locations
was similar to that of the halo dark matter, with the mean satellite location
measured with respect to the luminous hosts' major axes,
$\left< \phi_{\rm sat} \right>$, being only slightly less than the mean location of
the dark matter particles, $\left< \phi_{\rm dm} \right>$.
When AB06 allowed the degree
of alignment between
mass and light in the host galaxies to vary, the resulting locations of the satellites,
again measured with respect to the luminous hosts' major axes, varied from 
being nearly isotropic (caused by aligning the angular momenta of
the luminous hosts with the net angular momenta of the hosts' dark matter haloes)
to a complete reversal of the expected distribution, with the satellites showing
a preference for being located near the hosts' {\it minor} axes 
(caused by aligning the angular momenta of the luminous hosts with 
the largest principle axes of the hosts' dark matter haloes).

\'Ag\'ustsson \& Brainerd (2010; hereafter AB10) investigated the locations
of the satellites of relatively isolated host 
galaxies in a mock redshift survey of the first Millennium
simulation (MS; Springel et al.\ 2005).  The luminous galaxies in AB10
were obtained from the
DeLucia \& Blaizot (2007) SAM, which included rest-frame luminosities in 
multiple bandpasses and $B$-band bulge-to-disc ratios.  AB10
embedded the luminous host galaxies within their dark matter haloes in
various ways, and they explored the effects of different host galaxy orientations
on the resulting satellite locations.  
Using the $B$-band bulge-to-disc ratio, AB10 divided their
hosts into those with expected disc morphologies and
those with expected elliptical morphologies.  AB10 found that
the only way they could reproduce the observed dependence of SDSS satellite
locations on host colour, stellar mass, and star formation rate
was for the MS host galaxies to be embedded within their
dark matter haloes in markedly different ways: mass and light had to be
well-aligned in elliptical hosts, resulting from a model in which luminous
ellipticals were essentially miniature versions of their dark matter
haloes, and mass and light had to be poorly-aligned within disc hosts, resulting
from a model in which the angular momenta of the luminous discs 
were aligned with the net angular momenta of their dark matter
haloes.  Within a given host morphological class (i.e., elliptical vs.\
disc), AB10 found that the locations of the MS satellites were 
{\it independent} of the hosts'
stellar mass and star formation rate, and it was only by combining the 
locations of the satellites of both elliptical and disc MS hosts that the 
observed dependence of SDSS satellite locations on host colour, stellar mass
and star formation rate could be reproduced by the MS. 

More recently, Dong et al.\ (2014; hereafter Dong14) 
explored the locations of satellite galaxies
in a $\Lambda$CDM simulation that included smoothed-particle
hydrodynamics.  Unlike AB10, who had the liberty of choosing the 
ways in which the luminous
hosts were oriented within their dark matter haloes,
Dong14 were restricted to the use of
the luminous, stellar particles
to define the orientations of their hosts.
Dong14 used the stellar particles
to compute reduced inertia tensors,
the eigenvalues of which were used to 
define the hosts' luminous major and minor axes in the plane of the sky.  Dong14 
concluded that, on average, satellite galaxies were found preferentially
close to the major axes of their luminous hosts, with the degree of 
satellite anisotropy being most pronounced for satellites with the highest
metallicities.  Overall, Dong14 found that the mean satellite
location, measured with respect to the hosts' luminous major axes,
was essentially independent of satellite
stellar mass and was weakly-dependent on both the host stellar mass and 
the host
halo virial mass, with the satellites of the most massive hosts showing the
greatest degree of anisotropy.  Dong14 do not discuss the details of
the host and
satellite selection criteria for their sample, so it is unclear whether their
systems are directly comparable to those in AB10.  The minimum
satellite stellar mass in Dong14 is 
an roughly an order of magnitude larger than the minimum 
satellite stellar mass in AB10; however, the range of 
host stellar masses in Dong14 is similar to that of AB10.

In the case of relatively isolated hosts, studies of the 3-d locations 
of satellite galaxies obtained from SAMs 
are generally in agreement that the satellites trace the underlying 
dark matter mass
distribution (see, e.g., Sales et al.\ 2007; Wang et al.\
2014).  However, \'Ag\'ustsson \& Brainerd (2018) found that
the degree to which MS satellites traced the underlying mass distributions
of their hosts was a strong function of the physical properties of the
hosts, with the satellites of red hosts tracing the 3-d host mass
distribution well and the satellites of blue hosts having a 3-d spatial
distribution that was twice as concentrated as the hosts' dark matter mass.
In addition, a study of the 3-d spatial distributions of
satellite galaxies in the hydrodynamical Illustris-1 simulation 
by Brainerd (2018) showed
that the satellites do not trace the hosts' mass distributions
at all well on scales less than 50\% of the halo virial radius.   If it is
the case that satellite galaxies do not, in general, trace their
hosts' mass distributions,
then the interpretation of any observed spatial anisotropy 
becomes particularly complicated.

There are a number of reasons to expect that satellite distributions
obtained from SAMs and hydrodynamical simulations may differ from one another.
First, since SAMs are run in conjunction with `dark matter only' only simulations,
they do not provide luminous shapes for the host galaxies.  Therefore, to measure
satellite locations with respect to the orientations of luminous host
galaxies, assumptions must be made about the way in which the luminous hosts
are embedded within their dark matter haloes (e.g., AB10). Oftentimes the
assumptions that are made are overly simplified, despite being
physically motivated.  In the case of hydrodynamical
simulations, the stellar particles within the hosts allow for a direct determination
of the orientations of the host galaxies within their dark matter haloes, and these
orientations
may well differ from the assumptions adopted for the analysis of the host-satellite
samples obtained from SAMs.  Second, it has long been understood (e.g., Blumenthal
et al.\ 1986) that dark matter haloes react to the infall of baryonic material, resulting
in haloes that are {\it rounder} than if baryonic infall had not occurred.  Therefore,
we expect that, for a given host galaxy, its surrounding dark matter halo will have
a different shape in a hydrodynamical simulation than in a dark matter
only simulation.  If satellite galaxies are faithful tracers of the dark matter
surrounding the host galaxies, one would then expect the satellites in the hydrodynamical
simulation to have a distribution that is less isotropic than their SAM-derived
counterparts.

In this paper we use the high-resolution hydrodynamical
Illustris-1 simulation to obtain a
sample of relatively isolated host galaxies and their
satellites, and we further explore the locations 
of satellites with respect to their hosts.
Illustris-1 
is the highest resolution
simulation produced by the original Illustris Project (Vogelsberger et al. 2014a; Nelson
et al.\ 2015).  At the present epoch (i.e., $z = 0$), the gravitational force
softening length in Illustris-1 is 710~pc, the smallest hydrodynamical gas
cells are 48~pc in extent and the simulation contains
$\sim 40,000$ resolved, luminous galaxies.  
The simulation volume is a cubical box
with a comoving sidelength of $L = 106.5$~Mpc. Periodic boundary conditions,
$1820^3$ dark matter particles of mass $m_p = 6.3\times 10^6 M_\odot$, 
and $1820^3$ hydro cells with an initial baryonic mass resolution of
$1.3\times 10^6 M_\odot$ were used.  
Compared to the simulation used by Dong14,
the mass resolution of Illustris-1
is $\sim 100$ times greater and the force resolution is
$\sim 6$ times greater, allowing significantly better resolution of both
the luminous host galaxies and their satellites.
The values of the
cosmological parameters adopted for the
Illustris Project are: $\Omega_m = 0.2726$, $\Omega_\Lambda = 0.7274$, $\Omega_b = 0.0456$,
$\sigma_8 = 0.809$, $n_s = 0.963$, and $H_0 = 70.4$~km~s$^{-1}$~Mpc$^{-1}$.
All data used for our analysis are publicly available
through the Illustris Project website: {\tt http://www.illustris-project.org}.

The paper is organized as follows.  In \S2 we discuss the selection criteria for
our host-satellite sample and we summarise various sample properties.  In \S3
we use the full sample of
hosts and satellites to compute the locations of the satellites, in projection on the sky,
with respect to the major axes of the hosts' dark matter mass and the
hosts' stellar mass.  
In \S3 we also discuss the dependence of the 
satellite locations on
various physical properties of the hosts and satellites, the projected
separation between hosts and satellites, and on the definition of the 
hosts' stellar mass major axes.
In \S4 we compute the locations of the satellites from a subset of the
full host-satellite sample, where the properties of the subset
are similar to those of the host-satellite sample in AB10.
In \S5 we present a summary and discussion of our
major results.

\section{Host-Satellite Sample}

Observational studies of the locations of satellite galaxies have often adopted
selection criteria that were intended to yield a sample of large, bright host
galaxies that are relatively isolated from similarly large, bright galaxies.  These
relatively isolated host galaxies
are surrounded by 
a sample of smaller, fainter satellite galaxies that 
are located within a distance of $\sim 500$~kpc of their hosts (see, e.g., Sales \& Lambas
2004, 2009; Brainerd 2005; Azzaro et al.\ 2007; Bailin et al.\ 2008;
AB10; \'Ag\'ustsson \& Brainerd 2011).
In the observed universe, host-satellite samples are always 
contaminated to some degree by `interlopers' (i.e., false satellites whose 
locations on the sky,
apparent magnitudes and line of sight velocities relative to a particular host place them
within the satellite sample but they are, in fact, located too far from the host
to be genuine satellites).  In a simulation we have full
phase space information for all galaxies and, unlike observational studies that must rely
on redshift space selection, here we adopt selection criteria  in real space that 
incorporate the 3-d distances between the host galaxies and their satellites.  That is,
in our analysis of the locations of Illustris-1 satellites we focus only on those
objects which might be considered to be genuine satellites and our sample does
not include the interlopers that are inevitably found in observational catalogs.

In order to select hosts that are relatively isolated from similarly large, bright
galaxies, all hosts were required to be at least 2.5 times brighter than any other
galaxy found within a radius of 700~kpc.  In addition, because we wish to focus 
on host galaxies that are sufficiently massive
that they would have `regular' morphologies in the observed
universe, hosts were required to have stellar masses in the
range $10^{10} M_\odot \le M_\ast^{\rm host} \le 10^{12} M_\odot$.
Lastly, all host galaxies were required to be located at the centres of their
friends-of-friends haloes.  This final requirement
insured that the hosts were located at the centres of
the potential wells within which the satellites were moving.
Satellite galaxies were then defined to be all other galaxies
with absolute magnitudes $M_r < -14.5$ (i.e., comparable to 
the resolution limit for luminous galaxies at $z=0$; see, e.g., the
$z=0$ Illustris-1 galaxy luminosity function in Vogelsberger et al.\ 2014b),
located within a radius of 500~kpc of the host.

%
\begin{figure}
	\includegraphics[width=3.35in]{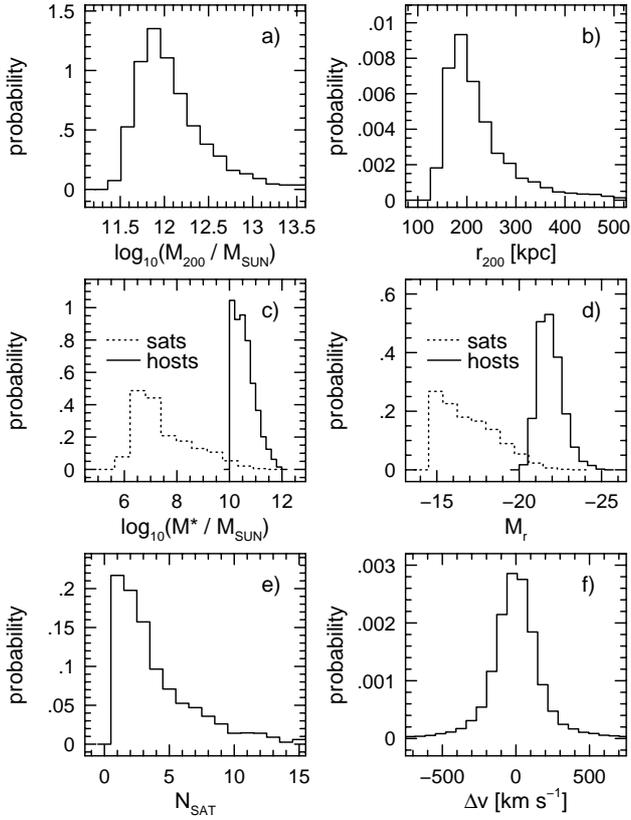}
\caption{Normalized probability distributions for various properties of the
host-satellite sample:
a) host virial mass, b) host virial radius, c) host and satellite stellar
mass, 
d) host and satellite absolute magnitude,
e) number of satellites per host, and f) line of sight
sight velocity difference between the hosts and satellites. 
}
    \label{fig:f1}
\end{figure}

Applying the above criteria to the $z=0$ timestep of Illustris-1 resulted
2,341 host galaxies with a total of 10,928 satellites.  
The median host virial mass is $M_{200} = 10^{12} M_\odot$
and the median host virial radius is $r_{200} = 205$~kpc.
The median host and satellite absolute magnitudes are 
$M_r^{\rm host} = -21.8$
and $M_r^{\rm sat} = -16.6$, the median host and satellite
stellar masses are $\log_{10}(M_\ast^{\rm host}/M_\odot) = 10.5$ and 
$\log_{10}(M_\ast^{\rm sat}/M_\odot) = 7.1$, and the median number of
satellites per host is $N_{\rm sat} = 3$.  
Figure~1 shows normalized
probability distributions for various properties of the host-satellite sample.
Compared to the host-satellite sample in AB10,
our sample here contains hosts with similar luminosities,
stellar masses and halo virial masses.  Our satellite sample extends to fainter
luminosities and smaller stellar masses than does AB10's sample. However,
and, because 
we did not impose a maximum line of sight velocity difference between our hosts
and satellites, the distribution of host-satellite velocity differences extends
somewhat beyond the value of $|dv|_{\rm max} = 500$~km/s imposed by AB10.  

\section{Satellite Locations: Complete Sample}

In analogy with observational studies, for which the satellite locations 
are determined in the plane of the sky, here we compute the locations of the 
Illustris-1 satellites
as viewed along each of the three principle axes of the simulation box.  Throughout,
we show results for the satellite locations computed as an average over these 
independent projections.  For a
particular host-satellite pair, the satellite location, $\phi$, is given by the angle between
the major axis of the surface mass density of the host and the direction vector that
connects the centroids of the host and satellite, as viewed in projection along a
given principle axis.  Since we are primarily interested in whether
the satellites show a preference for being aligned with either the major or minor
axes of their hosts, we restrict $\phi$ to the range $\left[ 0^\circ, 90^\circ \right]$,
where $\phi = 0^\circ$ indicates alignment with the host major axis,
$\phi = 90^\circ$ indicates alignment with the host minor axis, and a mean satellite
location of
$\left< \phi \right> = 45^\circ$, computed over many host-satellite pairs,
indicates a distribution that is isotropic on average.

We define each host's dark matter distribution 
to be the distribution of the dark matter particles within the host's virial radius
(i.e., the Illustris Project `Group\_M\_Crit200' data field). Further,
we define each host's stellar mass distribution to be the distribution of the stellar
particles contained within the radius at which the host's surface brightness profile
drops below 20.7~mag~arcsec$^{-2}$ in $K$-band (i.e., the Illustris Project 
`SubhaloStellarPhotometricsRad' data field).
The host galaxies are well resolved, allowing accurate determinations of the
relative orientations of their dark and luminous components.  The number of dark
matter particles in the hosts' dark matter haloes ranges from $2.0\times 10^3$ 
to $3.3 \times 10^7$ and the number of
stellar particles contained within the host galaxies themselves ranges from 
$7.0\times 10^3$ to $1.1\times 10^6$. 
We use the hosts' dark matter and stellar particles to compute mass weighted moments
of the dark matter and stellar surface mass densities as
$I_{xx} \equiv \sum_{i=1}^N m_i x_i^2$,
$I_{yy} \equiv \sum_{i=1}^N m_i y_i^2$, $I_{xy} \equiv \sum_{i=1}^N m_i x_i y_i$,
and we define the host ellipticity to be
$\epsilon = 1 - b/a$ where $a$ and $b$ are, respectively, the semi-major and 
semi-minor axis lengths.

\subsection{Locations Averaged over all Satellites}

We begin our investigation by computing
the mean satellite location using all host-satellite pairs.
For each host-satellite pair, 
we compute $\phi$ separately for two
different definitions of the host surface mass density: [1] the dark matter mass of the
host's halo (i.e., the dark matter mass contained within the halo virial radius) and
[2] the stellar mass of the luminous host galaxy.
Averaged over the entire sample of hosts and satellites, the mean satellite location
is $\left< \phi \right> = 38.1^\circ \pm 0.1^\circ$ when $\phi$ is computed using 
the major axes of the hosts' dark matter haloes and $\left< \phi \right> = 
42.8^\circ \pm 0.1^\circ$ when $\left< \phi \right>$ is computed using the major
axes of the hosts' stellar mass.  That is, as expected from previous
studies, when the locations of the satellites are averaged over the 
entire sample, the satellites have a preference for being located
near the major axes of their hosts, and the anisotropy is considerably more 
pronounced when $\phi$ is measured with respect to the major axes of the hosts'
dark matter haloes than when it is measured with respect to the hosts' luminous
mass.  

\begin{figure}
	\includegraphics[width=3.0in]{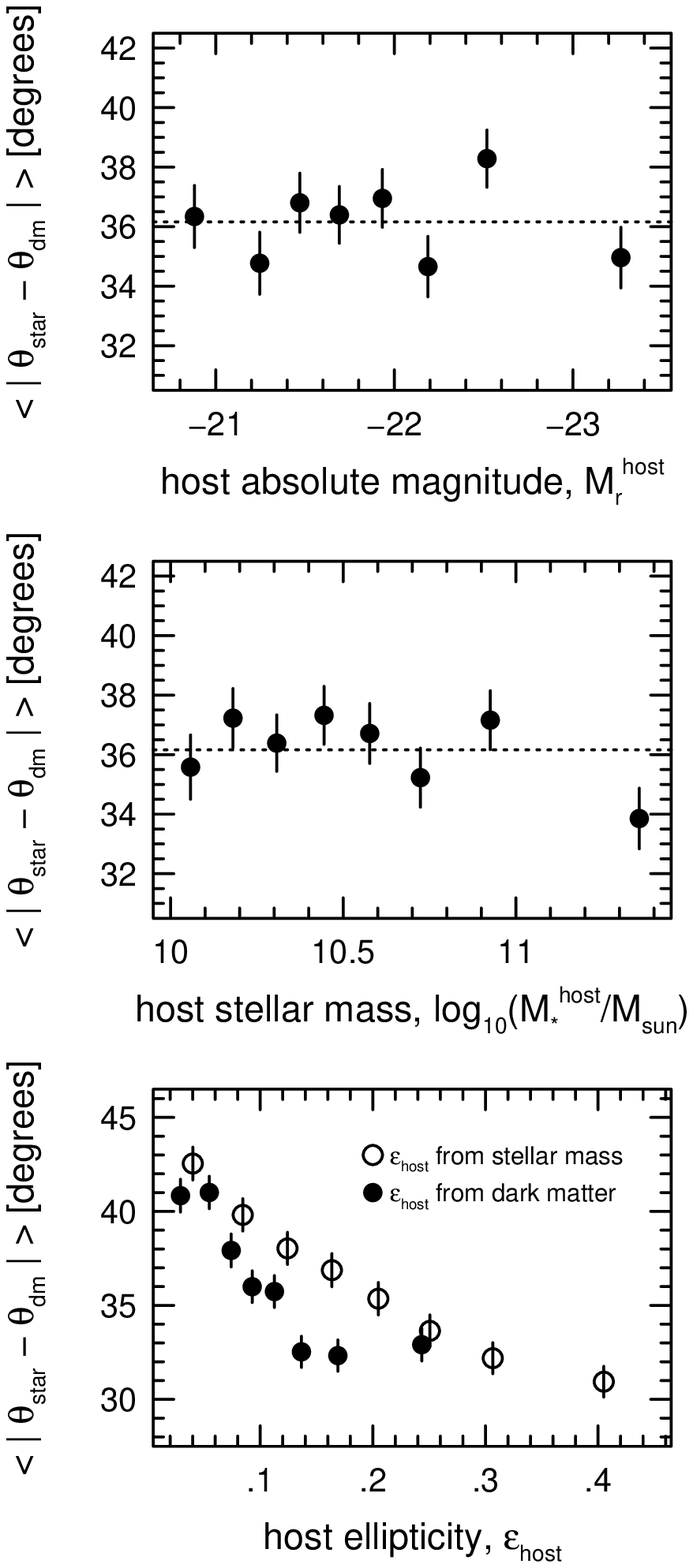}
\caption{Mean offset between the hosts' stellar surface mass distribution
and dark matter surface mass distribution,
$\left< | \theta_{\rm star} - \theta_{\rm dm} | \right>$.  
{\it Top:}
Dependence of
$\left< | \theta_{\rm star} - \theta_{\rm dm} |\right>$ on host
absolute magnitude.
{\it Middle:}
Dependence of
$\left< | \theta_{\rm star} - \theta_{\rm dm} |\right>$ on host
stellar mass.
{\it Bottom:}
Dependence of
$\left< | \theta_{\rm star} - \theta_{\rm dm} |\right>$ on host
ellipticity, for two different measures of the ellipticity:
stellar particles (open circles) and dark matter particles 
(filled circles). 
}
    \label{fig:f2}
\end{figure}

The reduction in the satellite
spatial anisotropy that
occurs when $\phi$ is computed using the hosts' stellar mass distribution is
due to the fact that the hosts' stellar mass distribution is not
perfectly aligned with the hosts' dark matter mass distribution.  Figure~2 shows
the mean offset between the position angle of the hosts' stellar mass
distribution, $\theta_{\rm star}$, and the position angle of the hosts'
dark matter mass distribution, $\theta_{\rm dm}$, as a function of host
$r$-band absolute magnitude (top panel), stellar mass (middle panel),
and ellipticity (bottom panel).  From the top and middle panels of Figure~2,
the mean offset between the hosts' dark matter surface mass density and 
stellar surface mass density is substantial, and is essentially
independent of host stellar mass and absolute magnitude.
This results in a significant reduction in the 
degree of anisotropy in the satellite locations when they are measured
relative to the hosts' stellar mass distribution.
The bottom panel of Figure~2 shows that 
the degree of offset between the hosts' stellar surface mass density and
dark matter surface mass density is a strong function of host ellipticity, independent
of whether the host ellipticity is defined to be that of
stellar mass or the dark matter halo.  
That is, the bottom panel of Figure~2 shows the mean offset between
the hosts' stellar mass and dark matter distributions is greatest for the
roundest hosts, but is significant even in the case of the 
flattest hosts.  Note that the data in Figure~2 are binned such that
there are a nearly identical number of objects per bin, resulting in 
similarly-sized error bars in each bin.

\subsection{Dependence of Locations on Host Properties}

Here we compute the mean locations of the satellites as a function of various
host properties.  In order to assess the degree to
which the satellites trace the host mass distribution in projection on the sky, 
we first compute 
$\left< \phi \right>$ for the satellites and for the mass particles as a function
of host ellipticity.  
The top panel of Figure~3 shows the mean satellite location, 
$\left< \phi_{\rm sat} \right>$, and the mean dark matter particle location,
$\left< \phi_{\rm dm} \right>$, as a function of the ellipticity of the hosts'
haloes.  In the top panel of Figure~3,
$\phi$ is computed relative to the major axes of the hosts' dark
matter surface mass density.  The bottom panel of Figure~3 shows
the mean satellite location, 
$\left< \phi_{\rm sat} \right>$, and the mass weighted mean stellar particle
location, $\left< \phi_{\rm star} \right>$, as a function of the 
ellipticity of the hosts' stellar mass distribution. In the
bottom panel of Figure~3, $\phi$ is computed relative to the major axes of
the hosts' stellar surface mass density.  From top panel of
Figure~3, then, the satellite
distribution is significantly flatter (i.e., considerably more anisotropic) than is
the dark matter distribution surrounding the host galaxies.  This result differs
from the results of AB06 who found that, in projection on the sky,
the satellite distribution was only slightly more anisotropic than the
dark matter surrounding the hosts (see the top panel of
Figure~3 in AB06). 
The bottom panel of Figure~3 shows that
for all but the very roundest hosts ($\epsilon_{\rm star} 
< 0.06$), the satellite distribution
is systematically rounder than is the hosts' luminous surface mass density.

\begin{figure}
	\includegraphics[width=3.0in]{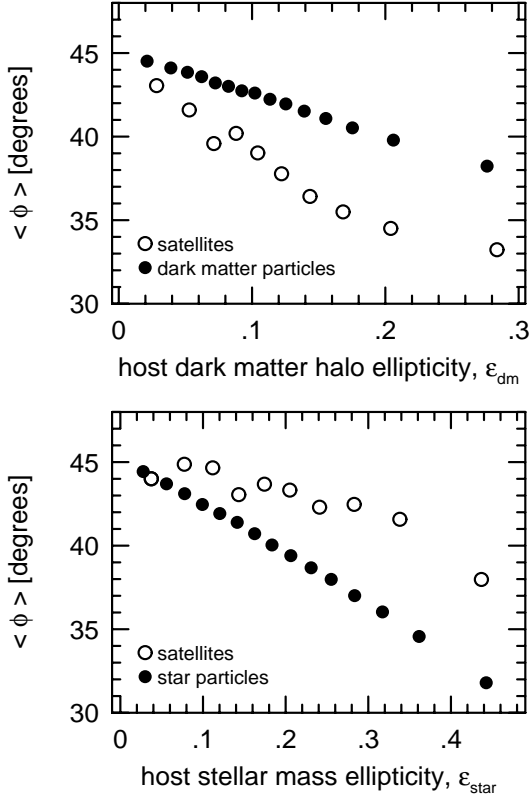}
\caption{Mean satellite location (open circles) and mass-weighted
mean particle location (filled circles) as a function of host
ellipticity.  {\it Top:} Location of satellites
and dark matter particles measured with respect to the major
axes of the hosts' dark matter surface mass densities; ellipticity
is that of the hosts' dark matter haloes.
{\it Bottom:} Location of satellites and stellar particles, measured
with respect to the major axes of the hosts' stellar surface
mass densities; ellipticity is that of the hosts' stellar mass.
Error bars are omitted since they are comparable to or smaller than
the data points.
}
    \label{fig:f3}
\end{figure}

Figure~4 shows the dependence of the satellite locations on the projected radius
between the hosts and satellites, $r_p$ (top panel), host absolute magnitude,
$M_r^{\rm host}$ (middle panel), 
and host stellar mass, $M_\ast^{\rm host}$ (bottom panel).
Filled circles indicate that $\phi$ was computed using the major
axes of the hosts' dark matter haloes; 
open circles indicate that $\phi$ was computed
using the major axes of the hosts' stellar mass distribution. 
The data in Figure~4 have been 
binned such that there are nearly identical numbers of host-satellite pairs
in each bin.

\begin{figure}
	\includegraphics[width=3.0in]{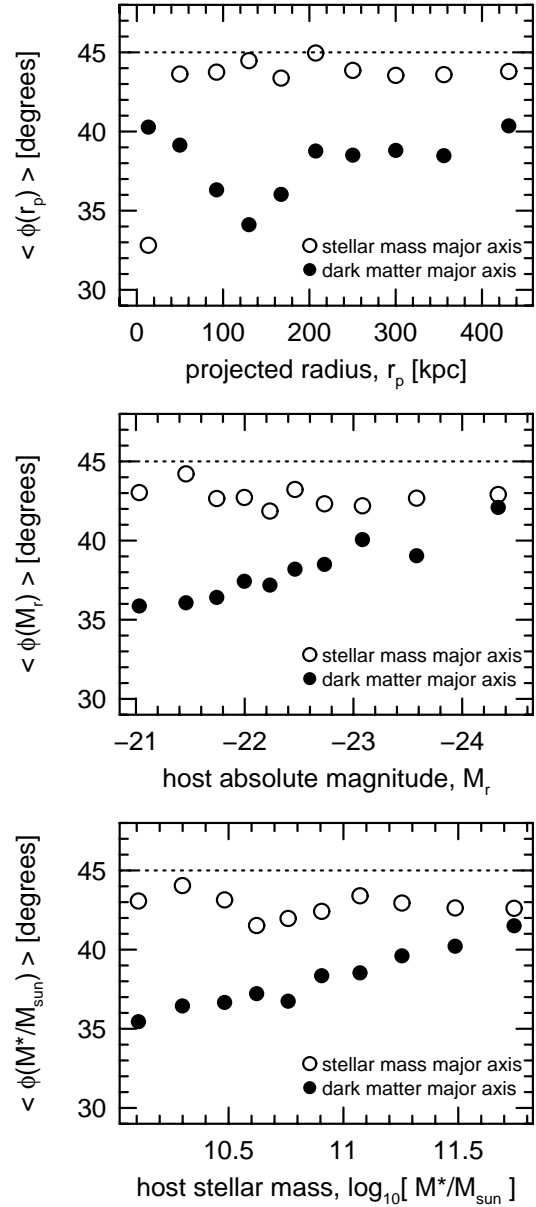}
\caption{Mean satellite location measured with respect to the
hosts' dark matter major axes (filled circles) and stellar mass
major axes (open circles).  {\it Top:} Mean satellite location as
a function of host-satellite projected radius.
{\it Middle:} Mean satellite location as a function of host
absolute $r$-band magnitude.
{\it Bottom:} Mean satellite location as a function of host
stellar mass.
Error bars are omitted since they are comparable to or 
smaller than the data points.
}
    \label{fig:f4}
\end{figure}

From Figure~4, there are significant differences between $\left< \phi \right>$ when 
measured relative to the hosts' dark matter surface mass density
vs.\ the hosts' stellar surface mass density.  
When computed relative to the hosts' stellar mass, 
$\left< \phi (r_p) \right>$ indicates that the satellites are distributed nearly
isotropically for 
$r_p \ga 30$~kpc, while on scales
$r_p \la 30$~kpc the anisotropy
is much more pronounced than when $\left< \phi(r_p) \right>$ is measured relative to
the major axes of the hosts' dark matter haloes.  
When measured with respect to the hosts' stellar surface mass density,
the difference in the degree of anisotropy in the satellite locations for
small and large values of $r_p$ is likely due to the locations of the
satellites in 3-dimensions.  In the top panel of
Figure~4, the three innermost bins for the results of $\left< \phi(r_p) \right>$,
measured relative to the hosts' stellar surface mass density, are centred on
projected radii $r_p = 13.4$~kpc, 49.8~kpc, and 92.4~kpc.  In these 
bins, the median 3-dimensional distances between hosts and satellites are,
respectively, 18.1~kpc, 82.0~kpc, and 127.2~kpc.  That is, within the bin centred 
on $r_p = 13.4$~kpc, the vast majority of the satellites are sufficiently close to
their host galaxies in 3-dimensions to be influenced by the baryonic mass of
the hosts.  In all of the other bins, the satellites are sufficiently distant
from their hosts in 3-dimensions that the influence of the baryonic mass of the
hosts is negligible.

When $\left< \phi(r_p) \right>$ is
computed relative to the major axes of the hosts' dark matter haloes, the degree of 
anisotropy increases with $r_P$ for projected radii less than 150~kpc, while for
$r_p > 150$~kpc the degree of anisotropy decreases with $r_p$.  Over all scales, however,
the sense of the satellite anisotropy remains the same: a preference for location near
the major axes of the hosts, independent of whether the major axis is that of the
dark matter mass or the stellar mass.  The results for $\left< \phi(r_p) \right>$
in the top panel of Figure~4 are in generally good agreement with the results
from AB10, who found that, when measured relative to the luminous major
axes of the hosts, $\left< \phi(r_p) \right>$ was only weakly dependent on 
$r_p$.  For $r_P < 50$~kpc, AB10 found very few satellites in the MS and, because
of this, it is not possible to compare our present results for 
$\left< \phi(r_p) \right>$ on scales less than 50~kpc to those of AB10.

Also from Figure~4, there are significant differences between the dependence of
the satellite locations on host absolute magnitude and stellar mass when
$\left< \phi \right>$ is measured relative to the hosts' dark matter mass vs.\
the hosts' stellar mass.  When measured relative to the major axes of the hosts'
stellar mass, $\left< \phi(M_r^{\rm host}) \right>$ and 
$\left< \phi(M_\ast^{\rm host}) \right>$ are
essentially independent of the physical properties of the hosts.  In contrast, when
$\left< \phi(M_r^{\rm host}) \right>$ and $\left< \phi(M_\ast^{\rm host}) \right>$ are measured with 
respect to the major axes of the hosts' dark matter haloes, both are monotonically
increasing functions. That is, the greater is $M_r^{\rm host}$ or
$M_\ast^{\rm host}$, the more isotropic is the satellite distribution.

Our results for 
$\left< \phi(M_\ast^{\rm host}) \right>$ in Figure~4 differ from the results of
AB10 and Dong14.  When measured relative to the hosts' luminous
major axes, both AB10 and Dong14 
found $\left< \phi(M_\ast^{\rm host}) \right>$ to be a monotonically
decreasing function of $M_\ast^{\rm host}$ (i.e., the
satellites of the most massive hosts exhibited a greater degree of
anisotropy in their locations than did the satellites of the least
massive hosts).  AB10 also found the dependence of the
satellite location on $M_\ast^{\rm host}$ to be considerably stronger
than did Dong14.  In addition, AB10 found that when the locations
of the satellites were measured relative to the hosts' dark matter major axes,
$\left< \phi(M_\ast^{\rm host}) \right>$ was {\it independent} of $M_\ast^{\rm host}$,
in contrast to the monotonically increasing function that we find in 
the bottom panel of Figure~4.

\subsection{Dependence of Locations on 3-d Host-Satellite Distance}

To address differences between 
previous results 
for the dependence of the mean satellite location
on host stellar mass and absolute magnitude and our own results,
we compute $\left<\phi \right>$
as a function of 3-d distances between the hosts and satellites.  We also
compute the dependence of the radial distribution of satellites on host
stellar mass and absolute magnitude.  The results are shown in Figure~5.
The top panel of Figure~5 shows that when the mean satellite location
is computed with respect to the hosts' dark matter major axes, $\left< \phi \right>$
is a strong function of the 3-d distance between the hosts
and satellites.  For satellites located within a distance 
$r_{3d} \la 180$~kpc, the mean satellite location decreases 
as a function of $r_{3d}$, indicating that, on average, the degree
of anisotropy in the satellite locations increases with $r_{3d}$
when $r_{3d} \la 180$~kpc.  For satellites with 3-d distances
$r \ga 200$~kpc (i.e., a distance comparable to the median
host virial radius), the mean satellite location increases as
a function of $r_{3d}$, indicating that, on average, the 
degree of anisotropy in the satellite locations decreases 
when  $r_{3d} \ga 200$~kpc.  This is unsurprising since 
the spatial distribution of satellites that are located outside their
hosts' virial radii should not reflect
the shapes of the hosts' dark matter haloes as well as those
that are found within the virial radii.

\begin{figure}
	\includegraphics[width=3.0in]{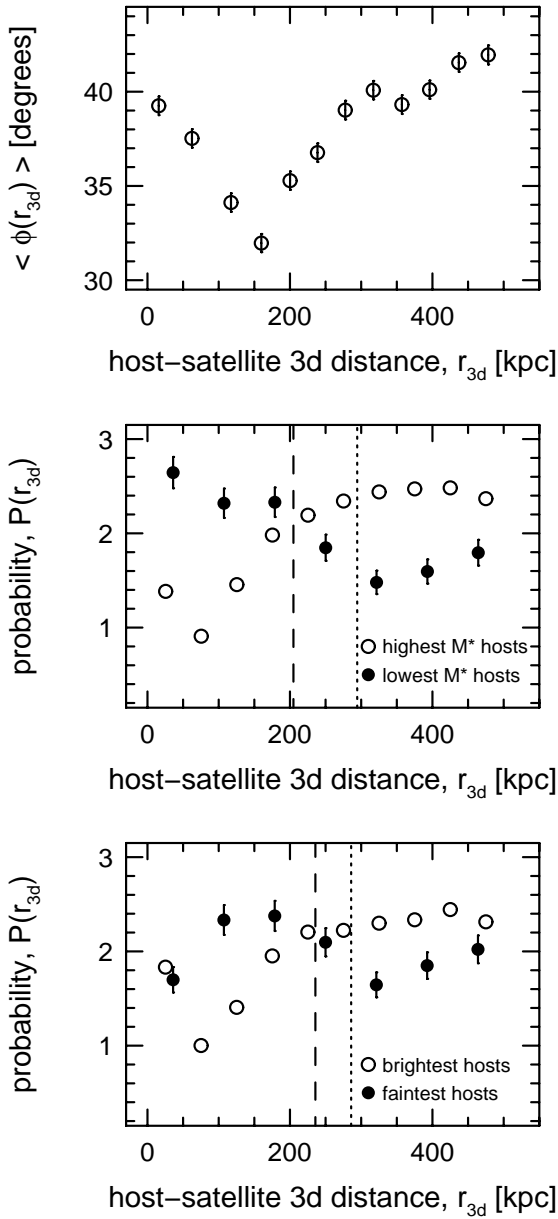}
\caption{
{\it Top:} Mean satellite location, measured with respect to the
hosts' dark matter major axes, as a function of 3-d host-satellite
separation, $r_{3d}$.
{\it Middle:} Probability distribution for 3-d host-satellite distances
as a function of $r_{3d}$ for hosts with 
$\log_{10}(M^\ast/M_\odot) < 10.24$ (least massive 25\% of the hosts;
filled circles) and
$\log_{10}(M^\ast/M_\odot) > 10.81$ (most massive 25\% of the hosts;
open circles).  Vertical lines indicate the mean virial radii of
the hosts in the two subsamples (dashed line: least massive hosts;
dotted line: most massive hosts).
{\it Bottom:} Probability distribution for 3-d host-satellite distances
as a function of $r_{3d}$ for hosts with $r$-band absolute magnitudes
$M_r > -21.36$ (faintest 25\% of the hosts; filled circles) and
$M_r < -22.34$ (brightest 25\% of the hosts; open circles).
Vertical lines indicate the mean virial radii of the
hosts in the two subsamples (dashed line: least luminous 
hosts; dotted line: most luminous hosts).
Error bars are omitted when they are comparable to or 
smaller than the data points.
}
    \label{fig:f5}
\end{figure}

The middle panel of Figure~5 shows probability distributions
for the host-satellite 3-d distances for the most massive
25\% of the
hosts ($\log_{10}(M^\ast/M_\odot) > 10.81$; open circles) 
and the least massive
the 25\% of the hosts ($\log_{10}(M^\ast/M_\odot) < 10.24$; filled
circles).  Vertical lines in the middle panel of Figure~5
indicate the mean virial radii of the two host samples.  From the 
middle panel of Figure~5,  in the case of the lowest
mass hosts, the number of satellites
at a given 3-d host-satellite distance
decreases out to distances of
$r_{3d} \sim 300$~kpc (i.e., larger than the mean virial radius),
beyond which it remains constant.  In the case of the
the highest mass hosts, 
the number of satellites at a given 3-d host-satellite
distance increases continuously for
$r_{3d} \ga 75$~kpc.  Because 
of this, the satellite sample for the lowest mass hosts is
dominated by satellites that are close to their hosts, while
the satellite sample for the highest mass hosts is
dominated by satellites that are far from their hosts.

Since, from the top panel of
Figure~5,  the mean satellite location is a strong
function of $r_{3d}$, with satellites at
$r_{3d} \la 180$~kpc showing a greater degree of anisotropy
on average than satellites at $r_{3d} \ga 200$~kpc, 
the dependence of $\left< \phi(M_{\rm host}^\ast/M_\odot) \right>$ 
in the bottom panel of Figure~4 is a reflection of the
3-d distribution of the satellites.  That is,
when measured relative to the dark matter major axes
of the hosts, the satellites of the least massive hosts
show a greater degree of anisotropy in their spatial
distribution than do the satellites of the most 
massive hosts (i.e., filled circles in the bottom panel
of Figure~4).  This is due to the fact that the satellite
population of the low-mass hosts is dominated by 
nearby satellites, with a high degree of anisotropy
in their spatial distribution, 
while the satellite population for
the high mass hosts is dominated by distant satellites,
with a low degree of anisotropy in their spatial 
distribution.

The bottom panel of Figure~5 shows probability distributions
for the host-satellite 3-d distances for the most luminous
25\% of the hosts and the least luminous 25\% of the hosts.
Since stellar mass and luminosity are strongly correlated,
the trends for the number of satellites as a function 
of $r_{3d}$ in the bottom panel of Figure~5 are similar
to those in the middle panel of Figure~5, albeit somewhat
weaker.  Overall, the bottom panel of Figure~5 indicates
that the satellite populations of the faintest hosts are
dominated by nearby satellites, and the satellite populations
of the brightest hosts are dominated by distant satellites.
Therefore, the dependence of 
$\left< \phi(M_r^{\rm host}) \right>$ in the middle
panel of Figure~4 is also a reflection of the fact that
the satellite population for low-luminosity hosts
is dominated by nearby satellites, while the satellite
population for high-luminosity hosts 
is dominated by distant satellites.

\subsection{Dependence of Locations on Satellite Properties}

Figure~6 shows the dependence of the satellite locations on the stellar masses
and absolute magnitudes of the satellite galaxies (bottom and top panels, 
respectively).
As expected from the misalignment of dark and luminous mass
in the host galaxies, the degree of satellite anisotropy is
significantly less when the satellite locations are computed relative to the 
major axes of the hosts' stellar mass than when they are computed
relative to major axes of of the hosts' dark matter mass.  In all cases,
$\left< \phi \right>$ decreases  with increasing satellite stellar mass and
luminosity.  That is, the most massive, most luminous satellites
show a greater degree of anisotropy in their locations than do the
least massive, least luminous satellites.  
Dong14 found that, when $\left< \phi \right>$ was
measured relative to the hosts' stellar mass, the satellite distribution
showed no dependence on satellite stellar mass.   The 
satellites in Dong14 were, however, considerably more massive than the majority
of our satellites.  If we consider only those satellites
in a similar stellar mass range, $10^9 M_\odot \la M_\ast^{\rm sat}
\la 10^{10} M_\odot$, our results agree with those of Dong14; the
locations of the satellites in this stellar mass range show no dependence
on stellar mass.
In contrast to our results and those of Dong14, AB10 found that,
when measured with respect to the luminous major axes of their hosts,
the locations of
MS satellites with stellar masses
$10^8 M_\odot \la M_\ast^{\rm sat} \la 10^{11} M_\odot$ showed a 
strong dependence on $M_\ast$, with the most massive satellites being
distributed much more anisotropically than the least massive satellites.

\begin{figure}
	\includegraphics[width=3.0in]{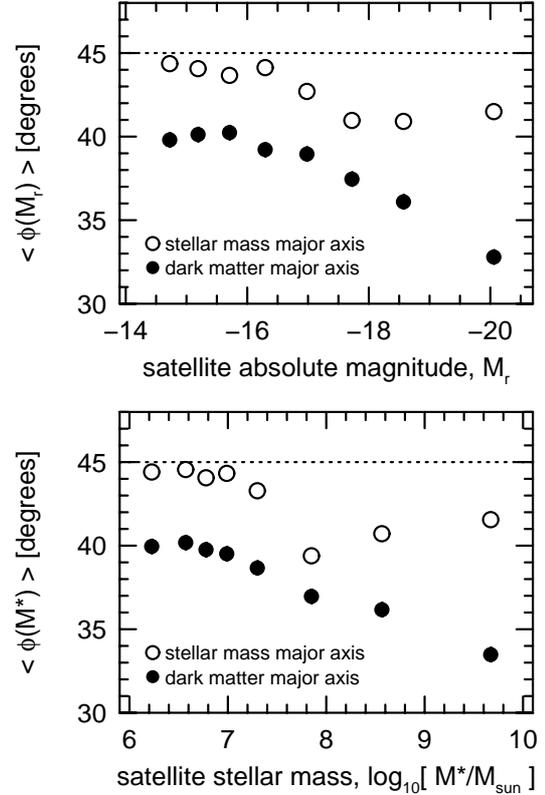}
\caption{Mean satellite location measured with respect to the
hosts' dark matter major axes (filled circles) and stellar mass
major axes (open circles).
{\it Top:} Mean satellite location as a function of satellite
absolute $r$-band magnitude.
{\it Bottom:} Mean satellite location as a function of satellite
stellar mass.
Error bars are omitted since they are comparable to or 
smaller than the data points.
}
    \label{fig:f6}
\end{figure}

\subsection{Dependence of Locations on Definition of Host Stellar Mass Major Axis}

In the previous sections, we defined the radius of the visible material in 
each host galaxy to be the radius at which the host's surface brightness
profile drops below 20.7 mag arcsec$^{-2}$ in $K$-band.  Within the Illustris
Project, this radius defines the total stellar mass for each host.
Since the orientation of a given luminous host
(i.e., the position angle of its stellar mass major axis)
may depend on the radius within which the orientation
is calculated,
here we explore the dependence of satellite locations on the way in which we
define the hosts' stellar mass major axes.  
We parametrise the radii we use in this section in terms of the fraction of the
total stellar mass contained within a particular radius, centred on each
host.  That is, we
define $M_\ast^{\rm host}(<R)$ as the amount of stellar mass 
contained within a projected
radius $R$ on the sky, centred on the host. 
We compute the hosts' stellar mass major axes using
various values of $R$, where in each case $R$ corresponds to 
some fixed fraction of the hosts'
total stellar masses, $M_\ast^{\rm host}(< R)/M_\ast^{\rm host}$. That is,
if $M_\ast^{\rm host}(< R)/M_\ast^{\rm host} = f$, $R$ corresponds to 
the projected radius within which a fraction $f$ of the total stellar mass
is contained. Below we allow $f$ to range from 0.1 to 1.0.

Figure~7 shows the ellipticities of the hosts' stellar 
surface mass density that we obtain using the total stellar mass of each host
(abscissa), together with the ellipticities we obtain when we use only a fraction of
the hosts' stellar mass (ordinate).  Figure~7 shows that, with the exception of
the very roundest hosts ($\epsilon_{\rm star}(R\ast) < 0.04$), computing
the hosts' ellipticity using a fraction $f < 1.0$ of the hosts' total stellar
mass leads to systematically smaller values of the host ellipticity (i.e., ``rounder'' 
hosts) than when the total 
stellar mass is used. This is likely a reflection of most of our hosts having
resolved central bulges (see, e.g., Snyder et al. 2015 for a discussion
of Illustris galaxy morphology at $z = 0$).  Given the significant
effect on the hosts' stellar mass ellipticities that we 
obtain when we use a fraction $f < 1.0$ of the hosts' stellar mass,
it is important to consider the degree to which our conclusions regarding
the locations of the satellite galaxies might be affected by the amount of stellar
mass we include in the calculation of the {\it orientations} of the hosts'
stellar mass major axes.

\begin{figure}
	\includegraphics[width=3.25in]{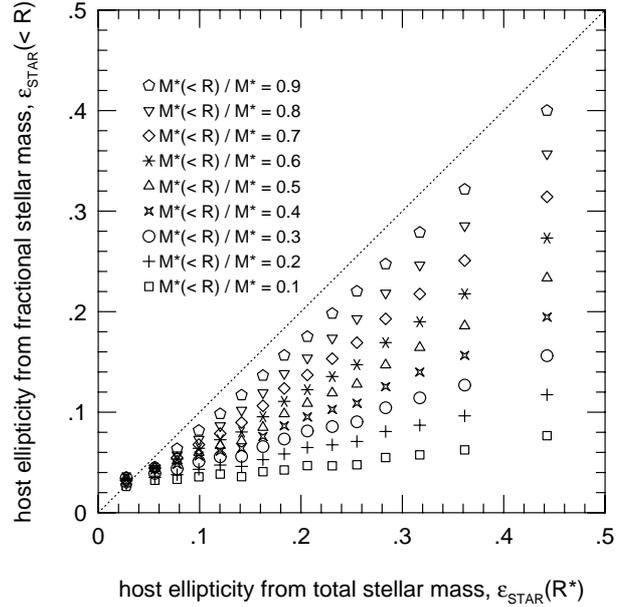}
\caption
{
Effect on host ellipticity, as measured from the stellar mass, that occurs
when the amount of stellar mass used in the calculation is varied.  {\it Abscissa:}
host ellipticity computed using the total stellar mass. 
{\it Ordinate:} mean host ellipticity obtained by using the stellar mass contained
within radii corresponding to fixed fractions, $f$, of the total stellar mass.
Error bars are omitted since they are smaller than the data points.
}
    \label{fig:f7}
\end{figure}

The top panel of Figure~8 shows the mean satellite location, computed
using all satellites, as a function of the amount of stellar mass within the hosts 
that we use to define their stellar mass major axes.  From
the top panel of Figure~8 we find that,
as long as the radius on the sky we use to define the 
hosts' major axes contains at least 70\% of the total stellar mass, there is 
no significant difference between $\left< \phi \right> $ computed using the total
host stellar mass and $\left< \phi \right> $ computed using
a fraction of the host stellar mass.  This
is reassuring because it suggests that the orientations of the hosts'
luminous major axes are
not especially sensitive to their outermost isophotes.  Hence, the mean locations
of the satellites are also not especially sensitive to the outermost isophotes
of their hosts.

The top panel of Figure~8 shows that even if we use $< 50$\% of the total
stellar mass to define the hosts' major axes, the satellites still show a significant
tendency to be found preferentially close to the hosts' major axes.  However, compared
to the mean satellite location we obtain using a host's total
stellar mass to define its major axis, the mean satellite
location is somewhat less anisotropic when we use only a fraction
of a host's stellar mass to define its major axes.  This is due to the fact that the
inner regions of the hosts are not perfectly aligned with the outer regions.
The bottom panel of Figure~8 shows the mean offset between the host major axes
computed using the total stellar mass and the host major axes computed using only
a fraction of the total stellar mass.  From the bottom panel of Figure~8, the smaller
the fraction of the stellar mass that is used to compute the hosts' major axes, the
greater is the offset from the orientation we obtain when we use the 
total stellar mass.  As long as at least 70\% of the total stellar mass is used to
compute the hosts' major axes, however, the mean offset between hosts' major axes computed 
using the total stellar mass and hosts' major axes computed 
using a fraction of the total stellar mass is 
$< 10^\circ$. This, then, leads to the agreement amongst the values of $\left< \phi \right>$
for $M_\ast^{\rm host}(< R) / M_\ast^{\rm host} \ge 0.7$ in the top panel of Figure~8.

Figure~9 shows the dependence of the mean satellite location on projected radius (top),
host absolute magnitude (middle), and host stellar mass (bottom) for three different 
definitions of the hosts' stellar mass major axes: 
$M_\ast^{\rm host}(< R) / M_\ast^{\rm host} = 1.0$ (open circles; identical to the open circles
in Figure~4),
$M_\ast^{\rm host}(< R)/ M_\ast^{\rm host}  = 0.75$ (triangles), and
$M_\ast^{\rm host}(< R)/ M_\ast^{\rm host}  = 0.5$ (squares).  
In almost every bin in Figure~9, the data points and error bars
are nearly identical to one another.  For legibility of the figure, therefore, we have
offset the data points corresponding to $M_\ast^{\rm host}(< R) / M_\ast^{\rm host} =0.5$
slightly to the left of those corresponding to 
$M_\ast^{\rm host}(<R) / M_\ast^{\rm host} = 1.0$.
Similarly, we have 
offset the data points corresponding to $M_\ast^{\rm host}(< R) / M_\ast^{\rm host} =0.75$
slightly to the right of those corresponding to 
$M_\ast^{\rm host}(<R) / M_\ast^{\rm host} = 1.0$.
From Figure~9, then, we find that as long as we include $\ge 50$\% of the stellar mass
of the hosts in the calculation of their major axes, the satellite locations
are insensitive to the amount of stellar mass that we use to define the hosts' major axes.

\begin{figure}
	\includegraphics[width=3.0in]{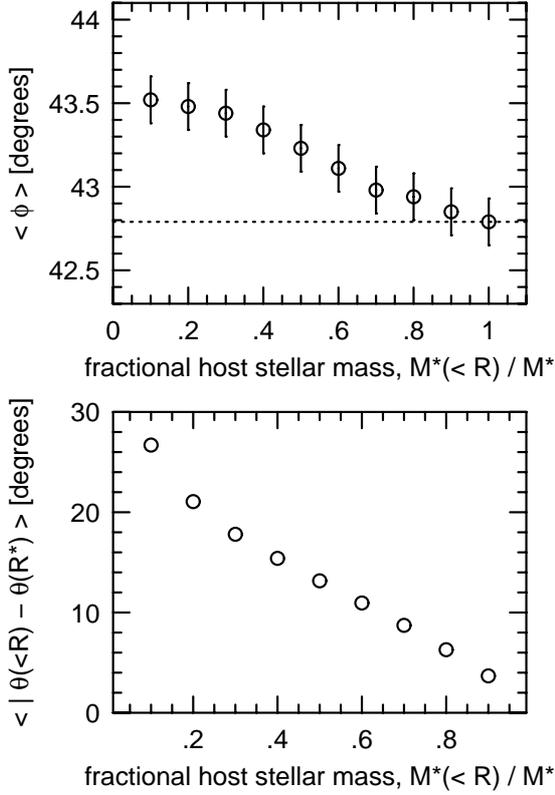}
\caption
{
{\it Top:} Mean satellite location, measured relative to the major axes
of the luminous hosts, as a function of the central fraction of the host's total
stellar mass 
used to define its luminous major axis.
{\it Bottom:} Mean offset, in projection on the sky, between the 
major axis of the host galaxy as defined using the total stellar mass and
the major axis of the host galaxy as defined using a central fraction of the
host's total stellar mass.
Error bars are omitted in the bottom panel because they are comparable to or smaller than
the data points.
}
    \label{fig:f8}
\end{figure}

\begin{figure}
	\includegraphics[width=3.0in]{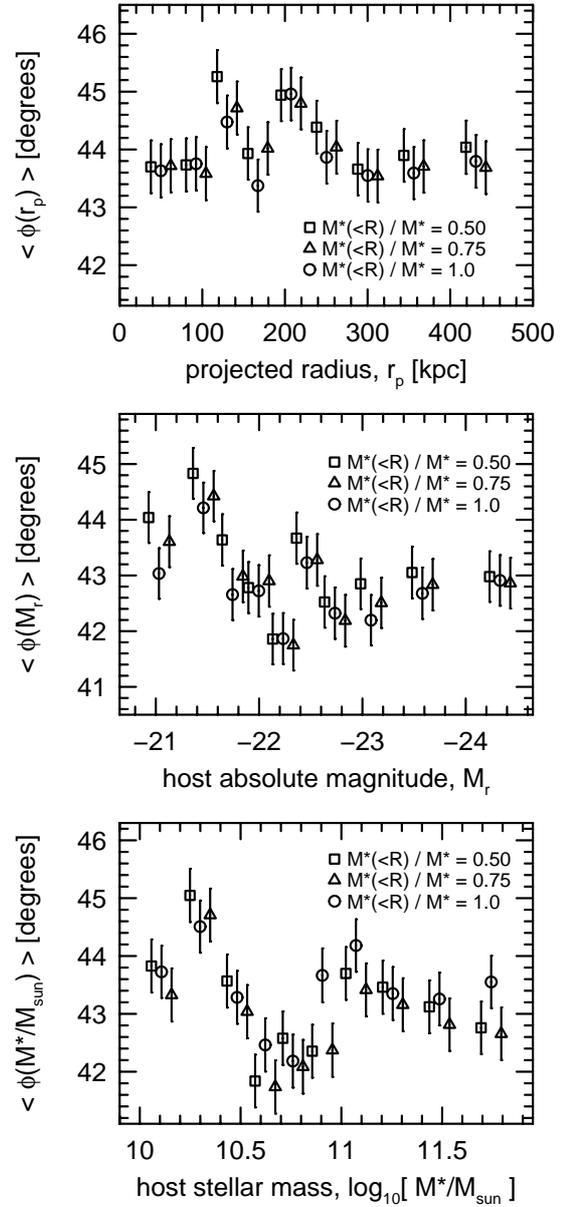}
\caption
{
Mean satellite location, measured relative to the hosts' stellar mass major axes,
as a function of the amount of stellar mass used to define the major axes.
{\it Circles:} Total host stellar mass is used to define
the hosts' major axes.  {\it Squares:} Innermost 50\% of the host stellar mass is used
to define the hosts' major axes.  {\it Triangles:} Innermost 75\% of the host stellar mass
is used to define the hosts' major axes.  Data points have been slightly
offset from one another in the horizontal direction to improve legibility of the figure.
{\it Top:} Dependence of satellite location on projected radius.
{\it Middle:} Dependence of satellite location on host $r$-band absolute magnitude.
{\it Bottom:} Dependence of satellite locations on total host stellar mass.
Note the considerably reduced vertical range in this figure as compared to Figure~4 and, hence, the
explicit inclusion of error bars here.
}
    \label{fig:f9}
\end{figure}

\section{Restricted Host-Satellite Sample}

Here our complete sample of hosts and satellites was selected 
in real 
space and incorporated 3-d distance information (see \S2).  In their study, 
AB10 selected host-satellite systems from a mock redshift survey of
the MS in order to better compare the results for MS satellites
to the observed locations of satellite
galaxies in the SDSS.  That is, AB10 selected their host and satellites
from the MS using redshift space criteria, rather than selecting them 
from real space.  Because of this, and because our complete sample of
satellites contains objects that were too faint to be resolved by the MS,
here we investigate the locations of Illustris-1 satellites in a
restricted sample that better matches the sample properties of AB10.
The properties of the restricted host-satellite sample are the following:

\begin{itemize}
\item $-24 \le M_r^{\rm host} \le -20$
\item $-22 \le M_r^{\rm sat} \le -17$ 
\item $M_r^{\rm sat} - M_r^{\rm host} \ge 2$
\item $10^8 M_\odot \le M_\ast^{\rm sat} \le 10^{11} M_\odot$
\item line of sight relative velocity between a host and its satellites
is $|dv| \le 500$km~s$^{-1}$
\item the projected radius between hosts and satellites is 
$r_p \ge 55$~kpc.
\end{itemize}

The results of measuring $\left< \phi \right>$ with respect to the
hosts' dark matter and stellar mass for the restricted sample are
shown in Figure~10.  From Figure~10, 
the dependence of the mean satellite
location on host-satellite projected radius, host absolute
magnitude, and host stellar mass in the restricted sample does
not differ considerably from the dependence shown by the complete
sample in Figure~4.  In particular, restricting our host-satellite
sample to one with properties similar to that of the sample in AB10
does not resolve the discrepancies between our results for the
locations of Illustris-1 satellites and those of AB10 for MS satellites.

\begin{figure}
	\includegraphics[width=3.0in]{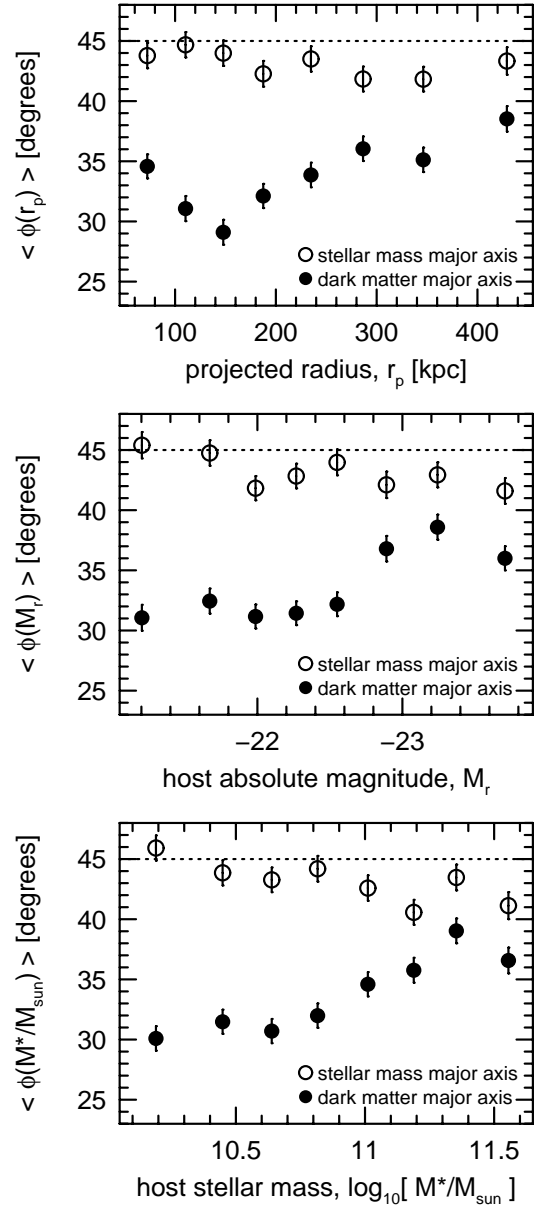}
\caption{Same as Figure~4 but for the restricted host-satellite
sample with properties that better match the MS host-satellite
sample in AB10.
}
    \label{fig:f10}
\end{figure}

\section{Summary and Discussion}

We have investigated the locations of luminous satellite galaxies in
the hydrodynamical Illustris-1 simulation.  The host galaxies were selected to
be relatively isolated and the mean satellite location,
$\left< \phi \right>$,  was computed
separately for two definitions of the host surface mass density: [1]
the dark matter mass of the host's halo 
and [2] the total stellar mass of 
the luminous host galaxy.  

In agreement with previous results, we find
that the satellites are distributed anisotropically in the plane of the
sky, with a preference for being located near the major axes of their 
hosts.  Due to misalignment between the dark matter surface mass density
and the stellar surface mass density, the degree of anisotropy in the
satellite distribution is significantly less when the mean satellite location
is measured with respect to the hosts' stellar surface mass density than
when it is measured with respect to the hosts' dark matter surface mass
density.  The mean misalignment between the host stellar surface mass density
and dark matter surface mass density is independent of host absolute
magnitude and stellar mass, but is strongly dependent on host ellipticity.
Overall, the satellite distribution is flatter in projection on the sky
than is the dark matter distribution surrounding the host galaxies.
With the exception of the roundest host galaxies, the
satellite distribution is rounder in projection on the sky than is the 
hosts' stellar mass distribution.

When measured with respect to the hosts' stellar surface mass
density, the mean satellite location is independent of host 
absolute magnitude and  host stellar mass.  This differs
from the results of AB10 and Dong14, both of whom found that the satellites
of the hosts with the largest stellar masses exhibited the
greatest degree of anisotropy.  
When measured with respect to the hosts' dark matter surface mass
density, the mean satellite location depends strongly on host absolute
magnitude and host stellar mass, with the satellites of the faintest,
least massive hosts showing the {\it greatest} degree of anisotropy. This
differs from the results of AB10, who found that the locations of
satellite galaxies in the MS were independent of host stellar mass
when the locations of the satellites were measured with respect to
the hosts' dark matter major axes.
We attribute this decrease in the anisotropy of the Illustris-1 satellite 
locations with host stellar mass and host luminosity to the 
fact that, for the brightest, most massive hosts, the
satellite sample is dominated by objects that are physically distant
from their hosts.  In the case of the faintest, least massive
hosts, the satellite sample is dominated by objects that are physically
close to their hosts.  Overall, the locations of satellites that 
are close to their hosts (e.g., within the virial radius) are much more
anisotropic than are the locations of satellites
that are far from their hosts.

We find that, as long as $\ge 70$\% of the hosts' total stellar mass
is included in the calculation of their stellar mass major axes,
the mean satellite location (computed as an average over
the complete sample) is insensitive to the amount of host 
stellar mass that is used to define the major axes.  We also
find that the smaller the fraction of the host total stellar mass we
use to define the major axes, the greater is the offset between
the mean host orientation computed using the total stellar mass
and the mean host orientation computed using a fraction of the total
stellar mass (i.e., the inner regions of the hosts are not
perfectly aligned with the outer regions). 
Despite this offset, the mean satellite location 
remains anisotropic, with a preference for the satellites to
be located near the major axes of their hosts.  However, the mean
value of the satellite location angle, $\left< \phi \right>$, 
indicates that the satellite locations are somewhat less anisotropic
when a smaller fraction of the host total stellar mass is used
to define the major axes.  In addition, we find that as long
as $\ge 50$\% of the hosts' stellar mass is used to define their
major axes, the dependence of the satellite locations on projected
radius, host absolute magnitude, and host stellar mass is 
unaffected by the amount of host stellar mass used to
define the major axes.  The
insensitivity of the satellite locations
to the amount of host stellar mass that is used to define
their major axes (provided a substantial amount of the stellar
mass is included) is reassuring because it indicates that the
satellite locations are not especially sensitive to the 
outermost isophotes of their host galaxies.  This is important because
in both `simulation space' and the observed universe, the outer regions
of galaxies blend smoothly into the overall mass distribution of the universe 
and the night sky background.

The locations of the 
brightest, most massive Illustris-1 satellites are
more anisotropic than are the locations of the faintest, least massive
Illustris-1 satellites.  When measured with respect to the host 
stellar surface mass density, the locations of
Illustris-1 satellites with stellar
masses in the range $10^8 M_\odot \le M_\ast^{\rm sat} \le
10^{10} M_\odot$ are independent of satellite stellar mass, in good
agreement with the results of AB10 and Dong14.

The cause of the differences between some of our results 
and those of AB10 and 
Dong14 are not entirely clear.  Restricting our host-satellite
sample to a sample with properties similar to that
in AB10 does not resolve the differences between our results
and theirs.
Importantly, however, AB10 made specific assumptions 
about the orientations of the luminous hosts within their dark
matter haloes since the galaxy catalogs in AB10 were based on 
the results of a SAM, not a hydrodynamical simulation.  Dong14
defined the orientation of the host stellar mass in their 
hydrodynamical simulation in the same way we have done here,
but it is not clear how Dong14 selected their host-satellite
systems. In particular, it is not clear whether their 
host-satellite systems are as
isolated as ours.
Given the different results for satellite locations found
by these studies, it will be interesting in future
to compare their results to similar analyses of other 
high-resolution cosmological simulations, including those
that incorporate SAMs and those that incorporate
numerical hydrodynamics.

\section*{Acknowledgements}

Insightful conversations with Patrick Koh 
and helpful suggestions from the anonymous referee
are gratefully acknowledged.  This work
was partially supported by Boston University's Undergraduate Research Opportunities
Program (UROP).





\begin{thebibliography}{99}


\bibitem[\protect\citeauthoryear{Agustsson}{2006}]{AB06}
\'Ag\'ustsson I., Brainerd T. G., 2006, ApJ, 650, 550 (AB06)

\bibitem[\protect\citeauthoryear{Agustsson}{2010}]{AB10}
\'Ag\'ustsson I., Brainerd T. G., 2010, ApJ, 709, 1321 (AB10)

\bibitem[\protect\citeauthoryear{Agustsson}{2011}]{AB11}
\'Ag\'ustsson I., Brainerd T. G., 2011, ISRAA, 2011, 958973

\bibitem[\protect\citeauthoryear{Agustsson}{2012}]{Ingi}
\'Ag\'ustsson I., 2012, PhD Dissertation, Boston University

\bibitem[\protect\citeauthoryear{Agustsson}{2018}]{AB18}
\'Ag\'ustsson I., Brainerd T. G., 2018, ApJ, 862, 169

\bibitem[\protect\citeauthoryear{Azzaro}{2007}]{APPZ}
Azzaro M., Patiri S. G., Prada F., Zentner A. R., 2007, MNRAS,
376, L43

\bibitem[\protect\citeauthoryear{Bailin}{2008}]{Bailin2008}
Bailin J., Power C., Norberg P., Zaritsky D., Gibson B. K.,
2008, MNRAS, 390, 1133
\bibitem[\protect\citeauthoryear{Blumenthal}{1986}]{Blumenthal1986}
Blumenthal, G. R., Faber, S. M., Flores, R., Primack, J. R. 1986,
ApJ, 301, 27
\bibitem[\protect\citeauthoryear{Brainerd}{2005}]{Brainerd2005}
Brainerd T. G., 2005, ApJ, 628, L101

\bibitem[\protect\citeauthoryear{Brainerd}{2018}]{Brainerd2018}
Brainerd T. G., 2018, 
ApJL, 868, L9

\bibitem[\protect\citeauthoryear{Colless}{2001}]{Colless2001}
Colless M., et al.\ 2001, MNRAS, 328, 1039

\bibitem[\protect\citeauthoryear{Colless}{2003}]{Colless2003}
Colless M., et al.\ 2003, preprint (arXiv:astro-ph/0306581)

\bibitem[\protect\citeauthoryear{DeLucia}{2007}]{DLB2007}
De Lucia G., Blaizot J., 2007, MNRAS, 375, 2

\bibitem[\protect\citeauthoryear{Dong}{2014}]{Dong2014}
Dong X. C., Lin W. P., Kang X., Wang Y. O., Dutton A. A.,
Macci\`o A. V., ApJ, 791, L33 (Dong14)

\bibitem[\protect\citeauthoryear{Fukugita}{1996}]{Fukugita1996}
Fukugita M., Ichikawa T., Gunn J. E., Doi M., Shimasaku K., 
Schneider D. P., 1996, AJ, 111, 1748

\bibitem[\protect\citeauthoryear{Hogg}{2001}]{Hogg2001}
Hogg D. W., Finkbeiner D. P., Schlegel D. J., Gunn J. E.,
2001, AJ, 122, 2129

\bibitem[\protect\citeauthoryear{Jing}{2002}]{JS02}
Jing Y. P., Suto Y., 2002, ApJ, 574, 538

\bibitem[\protect\citeauthoryear{Kauffmann}{1999}]{Kauffmann1999}
Kauffmann G., Colberg J., Diaferio A., White S. D. M., 1999,
MNRAS, 303, 188
\bibitem[\protect\citeauthoryear{Koch}{2006}]{Koch2006}
Koch, A., Grebel, E. K. 2006, AJ, 131, 1405
\bibitem[\protect\citeauthoryear{Metz}{2007}]{Metz2007}
Metz, M., Kroupa, P., Jerjen, H. 2007, MNRAS, 374, 1125
\bibitem[\protect\citeauthoryear{Metz}{2009}]{Metz2009}
Metz, M., Kroupa, P., Jerjen, H. 2009, MNRAS, 394, 2223
\bibitem[\protect\citeauthoryear{Nelson}{2015}]{Nelson2015}
Nelson D., et al.\ 2015, A\&C, 13, 12
\bibitem[\protect\citeauthoryear{Nierenberg}{2012}]{Nierenberg2012}
Nierenberg, A. M., Auger, M. W., Treu, T., Marshall, P. J., 
Fassnacht, C. D., Busha, M. T. 2012, ApJ, 752, 99
\bibitem[\protect\citeauthoryear{Sales}{2004}]{SL04}
Sales L., Lambas D. G., 2004, MNRAS, 348, 1236

\bibitem[\protect\citeauthoryear{Sales}{2007}]{SNLWC}
Sales, L. V., Navarro, J. F., Lambas, D. G., White,
S. D. M., Croton, D. J. 2007, MNRAS, 382, 1901

\bibitem[\protect\citeauthoryear{Sales}{2009}]{SL09}
Sales L., Lambas D. G., 2009, MNRAS, 395, 1184

\bibitem[\protect\citeauthoryear{Smith}{2002}]{Smith2002}
Smith J. A., et al., 2002, AJ, 123, 2121
\bibitem[\protect\citeauthoryear{Snyder}{2014}]{Snyder2014}
Snyder, G. F., et al. 2014, MNRAS, 454, 1886
\bibitem[\protect\citeauthoryear{Springel}{2005}]{Springel2005}
Springel V., et al., 2005, Nature, 435, 629

\bibitem[\protect\citeauthoryear{Strauss}{2002}]{Strauss2002}
Strauss M. A., et al., 2002, AJ, 124, 1810

\bibitem[\protect\citeauthoryear{Vogelsberger}{2014a}]{Vogelsberger2014a}
Vogelsberger M., et al., 2014a, Nature, 509, 177

\bibitem[\protect\citeauthoryear{Vogelsberger}{2014b}]{Vogelsberger2014b}
Vogelsberger M., et al., 2014b, MNRAS, 444, 1518

\bibitem[\protect\citeauthoryear{Wang}{2014}]{Wang2014}
Wang W., Sales L. V., Henriques B. M. B., White S. D. M., 2014,
MNRAS, 442, 1363
\bibitem[\protect\citeauthoryear{Yegorova}{2011}]{Yegorova2011}
Yegorova, I. A., Pizzela, A., Salucci, P. 2011, AA, 532, A105
\bibitem[\protect\citeauthoryear{York}{2000}]{York2000}
York D. G., et al., 2000, AJ, 120, 1579
\bibitem[\protect\citeauthoryear{Zentner}{2005}]{Zentner2005}
Zentner, A. R., Kravtsov, A. V., Gnedin, O. Y.,
Klypin, A. A. 2005, ApJ, 629, 219

\end{thebibliography}



\bsp	
\label{lastpage}
\end{document}